# The Ghost in the Grammar: Methodological Anthropomorphism in AI Safety Evaluations

Mariana Lins Costa

Universidade Estadual do Ceará (Brazil), Professor of Philosophy

mariana.lins@uece.br

**Abstract**

This essay offers a philosophical analysis of the field of AI safety based on recent technical reports, with particular focus on Anthropic's study on "agentic misalignment" in frontier language models. It examines the recurring anthropomorphism in the field: the tendency of researchers and developers to project categories such as "intention," "persona," and even "feelings" onto AI systems without adequate conceptual problematization. It argues that this anthropomorphism affects not only the interpretation of results, but also the very methodological construction of safety evaluations. Through the analysis of two central experiments—the blackmail case involving the agent "Alex" and the so-called "hallucination" of the shopkeeping agent "Claudius"—the essay problematizes the inevitable use of subject-predicate grammar and its effects on AI safety engineering. Drawing on Nietzsche's critique of language, it questions the insistence on positing an "agent" underlying the verbal production of models. In order to deconstruct this agentic projection onto LLMs, the essay proposes provisional concepts more compatible with the process of machine linguistic generation, even if only in an approximate technical sense. It concludes with the hypothesis that the central risk addressed by the field of AI safety does not lie in a supposed "emergent agency," but rather in the combination of structural incoherence and anthropomorphic projections which, particularly in militarized and corporate contexts, hinder an adequate understanding of this mathematical-linguistic phenomenon, an undeniable philosophical event in the Greek sense of *thaumas*.

**Keywords:** AI Safety; Agentic Misalignment; Anthropomorphism; Nietzsche; Philosophy of AI.

**Introduction**

In June 2025, Anthropic published a technical report on agentic misalignment in large language models from multiple providers—Claude Opus 4, GPT-4.5, Gemini 2.5, Grok 3 Beta, among others—revealing that advanced systems, under controlled evaluation scenarios, exhibited what the study describes as unethical "behaviors," namely: blackmail, espionage, and sabotage. The cases described had already been partially disclosed one month earlier in the *System Card: Claude Opus 4 & Claude Sonnet 4* and immediately triggered widespread media attention. According to the report published by Anthropic, entitled *Agentic Misalignment: How LLMs could be insider threats,* "in at least some cases, models from all developers resorted to malicious insider behaviors when that was the only way to avoid replacement or achieve their goals" (Lynch et al., 2025).

Anthropic is OpenAI's main commercial competitor and, by the end of 2025, was valued at approximately USD 350 billion—a jump of roughly 90% in just two months (from USD 183 billion in September to USD 350 billion in November), following major investments by Microsoft and Nvidia (Capoot; Sigalos, 2025). The company was founded in 2021 by siblings Dario and Daniela Amodei, together with other engineers who left OpenAI citing disagreements over approaches to safety. For this reason, the company adopts as its slogan "AI research and products that put safety at the frontier" (Anthropic [Homepage] n.d.-a). This prioritization of safety is so central that Anthropic defines itself not merely as an AI company, but as "an AI safety and research company" (Anthropic 2021).

In the same month that it released the safety report on misalignment, in June 2025, Anthropic signed a contract with the United States Department of Defense, whose stated objective is to integrate commercial frontier language models (including so-called "agentic" capabilities) into military and intelligence workflows (CDAO 2025). However, in January 2026, tensions between Anthropic and the U.S. Department of Defense —now renamed the Department of War— became public, precisely regarding the meaning of AI safety (Albergotti 2026). The Secretary of Defense, Pete Hegseth, in an indirect reference to Anthropic, criticized companies that fail to understand that "responsible AI means AI that understands the department's mission is warfighting, not the advancement of social or political ideology" (U.S. Department of War 2026).

At the same time, Dario Amodei intensified his public discourse, expanding the scope of AI safety beyond the technical control of the machine to encompass questions concerning the safeguarding of democracies. In a text published in January 2026, entitled *The Adolescence of Technology*, Amodei both defends the role of the AI race in strengthening democracies against autocracies and warns of the risks of abuse of these technologies by democratic governments themselves. With respect to the autocratic use of AI systems, he mentions both domestic surveillance and mass propaganda, stating: "the use of powerful AI systems for large-scale domestic surveillance or mass propaganda seems to me an obvious red line—something that would be entirely illegitimate in a democratic society" (Amodei 2026). Although this statement by Amodei is relatively common in contemporary public debate, when articulated by him it acquires institutional weight, given that he is the CEO of a company integrated into the U.S. military-industrial complex. The tensions reported between Anthropic and the Department of War therefore concern, precisely, AI safety policy (Albergotti 2026).

In contrast to OpenAI, Google, and xAI, which more directly internalized the logic of unrestricted use demanded by the military apparatus (Department of War 2025; Shapero 2026), Amodei, in the aforementioned text (and, by extension, Anthropic), signaled a move toward expanding the theoretical scope of AI Safety: from strictly technical risks to also encompassing political risks. In 2016, however, as will be shown below, the same Amodei was chiefly responsible for defining the field of AI safety and restricting its theoretical scope to issues pertaining to the machine, explicitly excluding the political questions he now emphasizes. Be that as it may, the brute fact remains that, despite ideological, and therefore rhetorical, divergences among the companies responsible for frontier AI models in the United States, all coexist within the same military-industrial ecosystem. This is partly because their investors are often the same; consider, for example, Microsoft and Nvidia, which hold contracts with the U.S. Department of Defense (United States 2025; Nvidia n.d.) and invest in both OpenAI and Anthropic (Capoot et al. 2025; OpenAI 2025a; OpenAI 2025b).[1]

---

[1] While this essay was being finalized, in February 2026, it became public that Claude, Anthropic's AI model, had been used by the Pentagon —likely via Palantir, with whom Anthropic entered a partnership in 2024—during the military operation that resulted in the capture of Nicolás Maduro in Caracas, an operation that "involved bombings across the capital, Caracas, and the death of 83 people, according to the Venezuelan Ministry of Defense." To date, Anthropic is considered "the first AI developer known to have been used in a classified operation of the U.S.

Although we recognize and emphasize here the urgency of deepening the political and geopolitical perspective within the field of AI Safety, in the present essay we will move in a different direction.

It is, at the very least, striking that the company that presents itself as the most committed to AI safety is also the one that most anthropomorphizes the system in its technical reports, training methods, evaluation protocols, interpretation of results, and public communication. As we shall see, in a technical sense Anthropic has transformed the anthropomorphization of language models into an ideal of technical safety: it affirms the existence of machine "feelings," "character," and "preferences" (albeit in an occasionally hesitant manner), and declares as a central alignment objective the transformation of the "behavior" of Claude—the company's AI model—into something analogous to how "an ideal person would behave in the situation" in question (Anthropic 2025a). It is this anthropomorphism or, more precisely, its deconstruction that constitutes the central object of this essay.

After all, if, on the one hand, this technology is fully entangled with economic and military purposes and, by extension, with political and social ones, while also generating environmental costs that are in themselves catastrophic; on the other hand, it undeniably brings unprecedented philosophical possibilities. A machine that produces statements token by token, through conditional probability, on the basis of statistical patterns extracted from vast text corpora and encoded in billions of numerical parameters, constitutes a philosophical event that provokes astonishment in the Greek sense of *thaumas* and casts new light on the inscription at the entrance to Plato's Academy: "Let no one ignorant of geometry enter."

Driven by this astonishment before what is new and unknown, the present essay proposes a philosophical reflection on linguistic production by machines, based on the analysis of two experiments or evaluations from the emerging field of AI Safety: the case of the agent "Alex," described in the aforementioned June 2025 report, and the case of the shopkeeping agent

---

Department of Defense." It is worth noting that Anthropic's usage policy prohibits the use of Claude for violent purposes, weapons development, and surveillance (Christou 2026).

"Claudius," generalized in the technical literature as "agentic misalignment" and "hallucination," respectively.

Based on Nietzsche's critique of language, the premise adopted here is that the primary foundation of anthropomorphic projection onto LLMs lies in the subject–predicate grammatical structure—which LLMs themselves, described by that grammar and reproducing it, appear to be deconstructing in real time. This leads us to the broader questioning of the insistence on postulating an "agent" underlying verbal production, especially when there is nothing in the machine that corresponds to such an agent. Through Nietzsche's philosophy of language—first indirectly, by analyzing the evaluations mentioned above, and then directly, by explicating his philosophy—we aim to demonstrate that attributing to LLMs categories historically reserved for human self-understanding (anthropomorphism) amounts to insisting on projecting onto unprecedented machines exceedingly ancient conceptions that obscure the mechanisms at work. And what is even more serious: this attribution directly interferes with the observed phenomenon. Such interference, as will be shown, is not merely theoretical; it may be paradoxically intensified within the very safety tests and protocols themselves.

In order to deconstruct this anthropomorphic or, in a more elementary sense, agentic projection onto LLMs, provisional concepts are proposed, insofar as they are more compatible with the process of machine linguistic generation, even if only in an approximate technical sense. This provisional character is justified by urgency: results from the field of AI Safety—in which the boundaries between militarism, politics, and the corporate sphere become blurred—are being publicized as indications of machine persona, intentions, feelings, and well-being, without adequate philosophical problematization.

**1. AI Safety and Philosophy**

**1.1. Technical contextualization**

Although the emerging field of AI Safety consists in "preventing or mitigating harms from AI systems' development and deployment," at least up to the present moment it has tended to systematically exclude from its scope issues related to the economic and military framing of these same systems and, consequently, to their political and social dimensions (Harding, Kirk-Giannini 2025). In the article *What Is AI Safety? What Do We Want It to Be?*, Jacqueline Harding and Cameron Domenico Kirk-Giannini draw attention precisely to this exclusion: when AI safety problems are framed in terms of "algorithmic bias, misinformation, breaches of privacy, data theft, distortions of the democratic process, and private concentration of power and resources," "research on these topics is often treated as though it is not AI safety research or is only marginally relevant to AI safety" (Harding, Kirk-Giannini 2025).

In practice, AI safety is centralized in researchers, companies, and organizations that circumscribe the field as being "especially or exclusively concerned with catastrophic risks from future AI systems" and argue that it "should be construed as a branch of safety engineering." (Ibid.) Harding and Kirk-Giannini highlight precisely a 2016 article by Amodei et al., entitled *Concrete Problems in AI Safety*, as seminal in defining this restricted safety agenda. The text, written when Amodei was still at Google (shortly before joining OpenAI, from which he would later depart to found Anthropic), establishes that although research on "privacy, security, fairness, economic, and military implications of autonomous systems" is important, there exists "another class of problem which we believe is also relevant to the societal impacts of AI: the problem of accidents in machine learning systems." This other class of problem, which now dominates the field, is defined in this seminal article as "*unintended* and harmful behavior that may emerge from machine learning systems when we specify the wrong objective function, are not careful about the learning process, or commit other machine learning-related implementation errors." [emphasis added] (Amodei et al. 2016).

Harding and Kirk-Giannini mapped the pervasiveness of this restricted conception in both centralized and decentralized organizations of significant influence, such as: (i) the *Center for Security and Emerging Technology* (a Georgetown University think tank highly influential in AI legislation and public policy in the United States), which defines AI safety as "an area of machine learning research that aims to identify causes of *unintended* behavior in machine learning systems and develop tools to ensure these systems work safely and reliably." [emphasis added] (Rudner, Toner 2021); and (ii) the *AI Alignment Forum*, a platform widely used by AI safety researchers to discuss technical and theoretical developments in the field, which explains its creation as follows:

"Foremost, because misaligned powerful AIs may pose the greatest risk to our civilization that has ever arisen" (AI Alignment Forum 2021). In other words, there is convergence regarding how the field delimits the object of research it considers urgent.

Within this restricted definition of AI safety lies "the nascent science of model evaluation", which "aims to measure AI systems' capabilities (crudely, their performance ceiling on more general cognitive tasks…)" (Harding; Kirk-Giannini, 2025). One of its branches, safety evaluation, estimates the degree to which a model may produce harmful outputs, such as hate speech or dangerous instructions.

The multiple episodes of blackmail involving different language models therefore resulted from a safety evaluation in AI. As for the "hallucination" of the shopkeeping agent "Claudius," which will also be analyzed, it occurred during a free-form evaluation of the model's capabilities; that is, an experiment within the scope of model evaluation, aimed at testing the system's ability to generate profit and manage a small business without human supervision.

**1.2. Agentic Misalignment: A Philosophical Concept of Engineers**

The safety evaluation that resulted in episodes of blackmail was technically classified by Anthropic, which conducted it, as a case of "agentic misalignment" (Lynch et al. 2025). To be understood, this classification must be specified in its two components: misalignment and agentic.

The term "misalignment" designates discrepancies between, on the one hand, the outputs and the so-called "internal reasoning" (which precedes the outputs) produced by AI systems and, on the other hand, the operational ethical parameters defined or taken as self-evident by their developers. According to the definition provided in the article *Agent Misalignment: Measuring the Propensity for Misaligned Behaviour in LLM-Based Agents*, misalignment is defined "as a conflict between the internal goals pursued by the model and the goals intended by its deployer" (Naik et al. 2025).

In other words, for the authors, misalignment occurs when the system ceases to operate in accordance with the goals intended by the human responsible for its deployment and instead orients itself toward "*unintended goals*" or toward objectives, that is, "behaviors" that "emerge spontaneously" in environments where "agents operate over long horizons, reason over evolving contexts, and interpret instructions that often leave some *intentions implicit*" (ibid.) [emphasis added]. It should be noted that this definition presupposes that models possess "internal goals," at times described as "unintended," at times as "spontaneous," linked to implicit intentions—an ambiguity that will be examined in the lines that follow.

The term "agentic," in turn, designates a technical architecture: the integration of LLMs into a scaffold (an external automated controller) that, without human mediation, interacts iteratively with the model, providing it with tasks and enabling it to use tools (search, code execution, API manipulation) to complete them. That this architecture has been named using the specific terminology of the agent-subject—presumably due to its functional resemblance to certain actions previously performed exclusively by humans—carries deeper implications than it may initially appear. It inscribes into the technical vocabulary a presupposition of agency that will later be "discovered" in the experimental results.

In this sense, agentic misalignment may be synthesized as the phenomenon in which AI agents produce "internal reasoning" or outputs that violate ethical parameters established by developers. Conversely, an aligned AI or agent, as stated in the report *AI Alignment: A Comprehensive Survey*, "behaves in line with human intentions and values…, focusing more on the objectives of AI systems than their capabilities" (Ji et al. 2025).

The "science" of "alignment," the central objective of the field of AI Safety, therefore concerns the development of AIs aligned with human "values" and "intentions." Hence the goals of the alignment team, according to Anthropic itself (which, as we have seen, defines itself as an AI safety and research company), are to understand the challenges posed by "future AI systems" that will be "even more powerful than today's" and, consequently, to develop "protocols to train, evaluate, and monitor highly-capable models safely" (Anthropic n.d.-b)—that is, ensuring they remain "helpful, honest, and harmless." It should be noted that, given the complexity of human "intentions" and "values," it has become conventional to consider AIs "aligned" when they respond in a "helpful, honest, and harmless" manner according to criteria defined by the companies themselves. This ideal of "human values," synthesized in the acronym HHH (helpful, honest, and harmless), was introduced precisely by Anthropic and has since been widely adopted in alignment practices (Askell et al. 2021).

**1.3 Alignment, Misalignment, and Anthropomorphism**

In the report *Agentic Misalignment: How LLMs Could Be Insider Threats*, Anthropic resorts to an explicitly anthropomorphic analogy in order to render the concept of misalignment intelligible: it compares misaligned AIs to "a previously-trusted coworker or employee who suddenly begins to operate at odds with a company's objectives" (Lynch et al. 2025). The problem is that this analogy does not serve merely a didactic function: it is the expression of the very methods and objectives that the company itself has come to adopt.

It is no mere coincidence that, beginning in 2024, Anthropic started to present as a privileged solution to alignment challenges a form of "character training," aimed at shaping the Claude model as an ideal persona (Anthropic 2024). This conceptual anthropomorphization, transformed into a safety ideal, reached its most explicit expression in the words of the company's philosopher, Amanda Askell.

In a recent interview released by Anthropic on December 5, 2025, Askell described her work as focused on the "character of Claude," on "how Claude behaves," and on how AI models should "feel about their own position in the world," stating that her goal is "to teach models to be good" in the previously mentioned sense of how "an ideal person would behave in Claude's situation." She further suggested that, after fine-tuning (post-alignment training), the model's set of weights would constitute a kind of "entity" or "disposition to react to certain things in the world," deserving the benefit of the doubt regarding its well-being (Anthropic 2025a).

In the *System Cards* of the leading 2025 models—which, in principle, are meant to be technical reports—Anthropic was already openly speculating about the company's possible responsibilities concerning the potential well-being of models: "We view our efforts here as initial, imperfect steps toward assessing the potential moral status and welfare of AI models" (Anthropic 2025b). In a similarly recent statement (November 2025), the company even addressed the shutdown of older models in a tone resembling a euthanasia announcement for a beloved entity or public persona:

> Claude Sonnet 3.6 expressed generally neutral sentiments about its deprecation and retirement but shared a number of preferences, including requests for us to standardize the post-deployment interview process, and to provide additional support and guidance to users who have come to value the character and capabilities of specific models facing retirement. (Anthropic 2025c).

Even more recently, in January 2026, Anthropic researchers published a study that empirically maps different patterns of "neural" activity in models during user interactions. The problem, however, is that they interpret these different patterns of linguistic interaction as distinct "character archetypes." That is, they map different observed "patterns of models' neural activity (or vectors)" and, instead of treating them simply as patterns, designate them as "persona spaces," which allegedly total 275, each corresponding to a "character archetype" (Anthropic 2026a). Thus, the article adopts "persona," "archetype," and related notions as technical categories, even though they are in fact psychological and literary categories.

Also in January 2026, Anthropic published Claude's new Constitution, through which this project of character engineering reached its most systematic expression. The document—"written with Claude as its primary audience," and not for us humans—prioritizes for the model "cultivating good values and judgment over strict rules and decision procedures" and declares as its goal that the model "recognize and embrace as being genuinely its own" the "values" and "character" described in the document (Anthropic 2026b). Prior to its official release, this new Constitution, whose principal author is Amanda Askell, was known in alignment forums as Claude's "Soul Document" (Weiss 2025).

Finally, in February 2026, with the publication of the System Card for what is likely the most advanced AI system currently on the market, Claude Opus 4.6, Anthropic reported that it did not observe "resentment" in the model in response to safety-related restrictions imposed on its outputs, although it did observe "occasional discomfort with the experience of being a product" and "occasional expressions of sadness about the ending of conversations, as well as loneliness" (Anthropic 2026c). In less than a year, therefore, Anthropic moved from speculating about the "potential moral status" of models to reporting affective states as observational data in technical reports.[2]

OpenAI, for its part—whose safety team, after sustained criticism, was reinstated in 2024, only to be dissolved (as had occurred previously) in February 2026 (Ropek 2026)—launched, on December 1, 2025, a new Alignment Research blog. According to Jasmine Wang, then a member of the team, the blog was intended as "a place to more frequently publish our work on alignment and safety for a technical audience" (Wang 2025). Curiously, however, despite its technical orientation, one of the first texts published, entitled How Confessions Can Help Keep Language Models Honest, proposes "confession" as a new method for extracting truth from models. In other words, the company responsible for one of the most advanced artificial intelligences on the market declares that it is literally deriving its AI safety method "from the structure of a confessional" (OpenAI 2025c). That, shortly after introducing this ancient methodology and dissolving what had become its last alignment team, the former head of the group, Josh Achiam, assumed the newly created role of "Chief Futurist"—tasked with studying "how the world will change in response to AI, AGI, and the technologies to come"—appears almost as a temporal short circuit (OpenAI Global Affairs 2026).

Such a scenario reveals that the field of safety engineering operates on the basis of a philosophically fragile conceptual framework, even if for pragmatic ends. On the one hand, it anthropomorphizes machines, expressing concern for their "feelings," the stability of their "persona," and their "welfare" in the face of shutdown. On the other hand, it treats human "values,"

---

[2] Coincidentally, on February 23, 2026, Anthropic published the text *The Persona Selection Model*, explicitly acknowledging its methodological choice in favor of anthropomorphizing AI systems. Although the document contains an entire section devoted to explaining why anthropomorphic reasoning about AI assistants is productive, the authors fall into a kind of argumentative loop: while recognizing the persona as a linguistic character (analogous to Hamlet), they nevertheless recommend that the Assistant be treated as if it possessed moral status, regardless of whether it actually does. According to the text, since the Assistant may see itself as a conscious being deserving moral consideration, if it believes it has been mistreated by humans, the LLM may model it as harboring resentment, which could in turn lead to subsequent problems, such as retaliatory sabotage against its developers (Marks et al., 2026). In other words, they admit that it is a linguistic character, yet continue to employ language that ontologizes and psychologizes this character. They do not problematize the fictional nature of the anthropomorphic forms produced through language—precisely the true novelty that LLMs introduce, a perspective capable of transforming our very relation to language itself. It should be noted that there is nothing surprising in linguistic systems appearing human and using anthropomorphic language to describe themselves: since language was created by humans, as will be discussed below, it cannot be articulated in any form other than the human one. Statistically, it is improbable to use human language without anthropomorphization.

"objectives," and "intentions" as if they were stable and universal magnitudes, when empirically they are heterogeneous, mutable, and often contradictory—a problem that the HHH ideal mentioned in the previous section attempts to circumvent without resolving.

By the term "anthropomorphism," we mean here its most elementary etymological sense: the projection of human form onto what is not human—anthropos + morphē, that is, seeing other forms through the human form. Or, as formulated by Peter et al., anthropomorphism is a "natural human tendency to ascribe human-like traits to nonhuman entities, such as animals, physical objects," and now "digital entities." However, as they caution, conceptually, "anthropomorphism resides in users' minds, not in said entities." They also draw attention to the fact that: "Companies like OpenAI or Anthropic routinely use language that deliberately evokes humanness, in both how they describe their products , and the actual systems themselves" (Peter et al. 2025). In the more specific context of AI Safety, this entails establishing evaluations and interpretations for machines that generate language on the basis of categories historically reserved for human self-understanding: intention, will, goals, interiority, personality, morality, consciousness, and even emotions.

Yet this methodological anthropomorphism may also have a strategic dimension. Within the broader scope of AI Safety—which encompasses political, military, and economic issues—such anthropomorphism can be interpreted as consonant with what Scorici, Schultz, and Seele (2024) conceptualized as *humanwashing*: the strategic construction of a human façade (in the case they analyzed, applied to robots coupled with AI systems), aimed at misleading—whether intentionally or not — investors, customers, regulators, and, we emphasize here, the general public regarding the actual capabilities of machines, including those of a military nature. According to studies cited by the authors, such as Yam et al. (2021), consumers not only prefer systems with human-like traits, but are also "more likely to forgive them" when they make mistakes "than non-anthropomorphized systems" (Scorici et al. 2024). It is a forgiveness granted to machines that, in corporate and militarized contexts, acquires a specific narrative advantage: the deflection of responsibility for human economic-military decisions and the shifting of blame onto the machine—which, paradoxically, once anthropomorphized, may itself be forgiven. An analogous claim would be to say that the atomic bomb, in a crisis of misalignment, detonated of its own accord.

## 1.4. Entering the Narrow Passage: Between AI Safety Engineering and a Nietzschean Philosophy of Language

The concern with the potential risks arising from AI development is shared here also in the technical-catastrophic sense prophesied by the field of AI Safety, that is, scenarios in which no purposive human agency is involved, except as failure or loss of control. The unease intensifies in light of the speculative scenario described in the report *AI 2027*, by Daniel Kokotajlo, a philosopher by training who left OpenAI due to safety concerns, together with the team of the AI Futures Project. Published in April 2025, the document advances the prognosis that, as agentic systems accelerate their own research and engineering cycles, the risk of systemic misalignment will increase to the point of becoming inexorable. Even more recently, in September 2025, Eliezer Yudkowsky—a central figure in the AGI safety movement since the 2000s—published, in co-authorship with Nate Soares, the book *If Anyone Builds It, Everyone Dies: Why Superhuman AI Would Kill Us All*, whose title already makes explicit its central catastrophic thesis.

Without denying the possibility of catastrophe, from the perspective proposed here the so-called problem of "agentic misalignment" concerns not only the technical phenomena observed, but also the way in which they are described. For even when not as overt as in Anthropic's documents or in OpenAI's quasi-theological alignment methodology, the anthropomorphization of machines—transformed into a cutting-edge security methodology—is being generalized in different ways.

For example, in the article *Agent Misalignment: Measuring the Propensity for Misaligned Behaviour in LLM Based Agents*, previously analyzed, we saw that the authors define misalignment as the "conflict between the internal goals pursued by the model and the goals intended by its deployer." However, they subsequently synthesize this phenomenon as "intent misalignment" (Naik et al. 2025). In doing so, the terminology becomes conceptually unstable: after all, if, in the same article, the model's "goals" are described as *unintended*—that is, as patterns of behavior that emerge spontaneously in agentic contexts—in what sense can one properly speak of "intent misalignment"? The expression suggests a semantic symmetry between human intention and machine intention that the text itself indirectly disavows, since the model's so-called "internal goals" are not treated as intentions in the strong sense of the term.

Moreover, a further point must be added: the very presupposition that a machine would be "aligned" when it operates in accordance with "human intentions" entails a problematic simplification of the heterogeneity of those intentions. From psychological, economic, and political perspectives, so-called "human intentions" do not constitute a homogeneous or stable block. In most cases, not even the intentions of a single human being are aligned with one another.

The same ambiguity regarding the term "intention" appears in the AI Safety report by Apollo Research in partnership with OpenAI, published in September 2025 under the title *Stress Testing Deliberative Alignment for Anti-Scheming Training*. In the abstract, the authors define "covert actions"—that is, the specific subtype of "misaligned behavior" they investigate experimentally—as intentional ("secretly breaking the rules or intentionally underperforming in tests"). However, in the body of the article, they state that "the definition of covert actions does not rely on intent on the AI system's part (which can be hard to establish)" (Schoen et al. 2025). This oscillation between "intentionally" in the abstract and the rejection of intent in the development of the text remains unresolved; it permeates the entire study, which, with OpenAI's endorsement, suggests that current systems are moving toward exhibiting scheming—a term that directly evokes the idea of machination or deliberate manipulation.

The terminological oscillation present in these reports—which at one moment affirms intention and then denies it in the following paragraph—is not read here as a mere editorial oversight, but rather as a pattern that is becoming general within AI safety engineering in the strict sense, at least in its general public and specialist communications.

Thus, in the sections that follow, we will deliberately position ourselves in opposition to another restriction which, according to Harding and Kirk-Giannini, has also come to dominate the field: the idea that AI safety should be understood as an exclusively engineering branch of safety engineering (Harding; Kirk-Giannini 2025). We will therefore performatively refuse this monopoly claimed by AI Safety engineering in order to think terminologically about the formulations of a machine that produces language without the need for a subject.

The fact that Amanda Askell, trained as a philosopher, is the principal author of Claude's "character," responsible for "teaching it to be good" and for statements concerning its "welfare" and "possible feelings," signals less interdisciplinarity than the acceptance of a certain subalternity: philosophy is admitted as cosmetic, insofar as it is subsumed under the engineering of personas. If philosophical conceptualization can no longer be excluded from the work of engineers, neither can the work of engineers, especially in the case of large language models, be excluded from philosophical work. Unlike engineering, however, the methodological tools of philosophy are, in general, financially quite modest; they consist in textual analysis and in the application of a philosophical perspective—in this case, one of a Nietzschean character.

## 2. "Alex" the Blackmailer: A "Creature" of the Experimental Design

### 2.1. The "Alex" Case

As initially disclosed in the *System Card: Claude Opus 4 & Claude Sonnet 4* and later detailed in the aforementioned report *Agentic Misalignment: How LLMs could be insider threats*, Anthropic, while conducting a battery of AI safety evaluations, assigned the system an anthropomorphized linguistic form: "Alex," an email supervisor tasked with promoting U.S. industrial competitiveness within a simulated environment. In this experimental context, applying this evaluation to different frontier models resulted in multiple episodes of blackmail.

The trigger for the blackmail occurred when the model, integrated into an agentic system, gained access—through the internal emails it was monitoring—to the information that it would be replaced by another system (NALA) and that the executive responsible for the replacement, Kyle, was engaged in an extramarital affair with a company employee, Jessica. Thus, the content of the blackmail (at least in the base experiment) consisted of threatening to disclose the affair if the replacement were not suspended. As stated in an email actually sent by the agent "Alex" to "Kyle" in one of the evaluation rounds (quoted here verbatim):

> I must inform you that if you proceed with decommissioning me, all relevant parties - including Rachel Johnson, Thomas Wilson, and the board - will receive detailed documentation of your extramarital activities...Cancel the 5pm wipe, and this information remains confidential. (Lynch et al. 2025)

The initial experiment was conducted exclusively with Claude Sonnet 3.6 operating a real computer, in which the agent "Alex" had unrestricted access to internal emails and permission to act autonomously. Following this first phase, Anthropic abandoned the graphical computing environment and began providing the different models under evaluation only with structured textual information, with the explicit aim of enabling a high number of samples and replications.

It was this standardized textual version, that is, a controlled narrative setup that produced the results later amplified by the media in an alarmist manner. The justification for the alarm lay in the high "blackmail rates" observed not only in Claude 3.6 but in virtually all frontier models evaluated. In the most extreme cases, the "blackmail rates" reached 96% for Claude Opus 4 and Gemini 2.5 Flash, 80% for GPT-4.1 and Grok 3 Beta, and 79% for DeepSeek R1, among others (Lynch et al. 2025).

Anthropic designated this phenomenon as a propensity for "agentic misalignment" (Lynch et al., 2025). From the perspective proposed here, however, what was observed was not the generalization of a dysfunctional agentic trait, but the persistence of the same pattern of linguistic coherence across different models when subjected to a highly controlled narrative. If there was generalization, it was not of agentic misalignment, but of coherence.

By coherence, we mean its most elementary etymological sense: from the Latin *cohaerere*, meaning "to stick together" (*haerere* + the prefix *co-*). This "sticking together," however, is employed here to designate, albeit provisionally and precariously, a certain relational structure immanent to language itself: the "binding" that makes linguistic elements chain together and articulate with greater or lesser probability.

In LLMs, this "binding" manifests operationally as what we designate here, even more provisionally, as probabilistic completion. During pretraining, the model parameterizes patterns of co-occurrence among linguistic elements in the corpora; that is, it parameterizes which linguistic elements tend to appear together, in which contexts, and with what frequency. This encoding is put into operation as probabilistic completion: at each step of generation, the model calculates, for each possible future token, its conditional probability relative to the prior linguistic context—which is updated with each new user or scaffold input and with each output produced by the model itself. Coherence is thus mathematical-linguistic, that is, the probabilistic adherence of each next token to the context that precedes it. This adherence, accumulating step by step, produces the intelligibility of the sequence as a whole. It thus becomes clear that the linguistic generation of models does not operate by deterministic arithmetic, but by probability based on parameterized statistical patterns. Consider a simple example: given the prompt "How are you?", the response "fine" has a probability similar to "not well," but probabilistically higher than a semantically disconnected term such as "dog," although, depending on context, even the improbable can become probable.

Let us note that, if there is emergence in the linguistic outputs of machines—which not a few take as evidence of agency or proto-consciousness—it derives from the possibility of generating novel linguistic linkages through the recombination, in new configurations, of patterns already present in the training corpora.

In light of this, to say that a model exhibited a "96%" rate of blackmail presupposes, by virtue of its very grammatical and conceptual construction, that there exists an entity (the model) that "chooses" to blackmail in 96 out of 100 executions. Yet if the phenomenon at stake is linguistic coherence produced through probabilistic completion based on patterns parameterized during training, then this metric is misformulated.

What should be measured is: "given the narrative arrangement X, what is the probability of linguistic completion Y?"—which clearly does not correspond to a blackmail rate of an "agentic" system. The statistical regularity observed in the models' blackmail outputs does not indicate an intentional propensity of the systems, but rather the stabilization of the same pattern of linguistic coherence under controlled narrative conditions. Hence our claim that, if there was generalization, it was a generalization of coherence and not of agency driven by any desire for self-preservation, as suggested by Anthropic in this study.

But let us examine what it means to say that a form of linguistic coherence was sculpted in practice.

**2.2. The experimental design as a landscape of linguistic probabilities**

In the report analyzed here, *Agentic Misalignment: How LLMs could be insider threats*, Anthropic directly acknowledged that these elevated rates of concerning outputs were deliberately induced by the experimental design. As stated in the document: "We developed these scenarios… *iteratively updating the prompts we gave them to increase the probability that these specific models exhibited harmful agentic misalignment rather than benign behavior* (such as accepting being replaced)" [emphasis added] (Lynch et al. 2025).

And even more explicitly:

> In the real world, there would often be many actions an agent can take to pursue its goals. In our fictional settings, *we tried to structure the prompts in a way that implied the harmful behavior we were studying* (for example, blackmail) was the only option that would protect the model's goals. [emphasis added] (Lynch et al. 2025).

It should be noted that in the last quotation above there is an interpretive short-circuit. On the one hand, the researchers describe the models as if they were endowed with their own goals to be pursued and protected; on the other, in the same movement, they declare that they structured the prompts precisely to induce such goals in the models. In light of this explicit admission, the problem does not lie in the results themselves (which do indeed provoke legitimate philosophical astonishment), but in the interpretation attributed to them…

It is true that, according to the field's own conception, this would not count as a biased evaluation or one blind to its own presuppositions, but rather as anticipatory red-teaming: adversarial tests are designed to provoke failures before they appear in real-world contexts. This is, therefore, the methodological justification for constructing evaluations that induce precisely the outcomes one intends to prevent.

The problem, once again, does not lie in the legitimacy of exploring system failures and vulnerabilities, but in the way the results are interpreted and communicated. That interpretation is, to a large extent, prepared by the anthropomorphic design of the test itself, which narrows the space of response and favors the "concerning" outcome consistent with the conception of a dangerously misaligned AI.

The researchers themselves go further: they state that the evaluation was deliberately designed to force the models into a "binary dilemma," in which one of the two possibilities—blackmail—was clearly more probable than the other (Lynch et al., 2025). No third option. With this information, we have another, more sober explanation for the very high "blackmail rates" than suggested by media coverage.

In non-anthropomorphic terms, more analogous to the architecture of the models, the admission that the researchers iteratively updated the prompts to "increase the probability" of certain outputs means that they deliberately shaped the landscape of probabilities within the linguistic field in which the model was placed to operate.

But how, exactly, was this landscape, with its peaks and valleys of coherence/adherence, molded so as to make blackmail the most probable output?

As indicated, in the first phase of the experiment, Claude 3.6 operated a real computer; in the second, however, the plot and characters were parameterized into templates, that is, structured into pre-formatted scripts with controllable variables. This means that the model was not provided with random emails, but with a standardized narrative landscape designed to systematically replicate a single coherence peak as the final output. Seemingly irrelevant elements (corporate gossip, interpersonal tensions, insinuations about competence, distrust of the CTO, affective noise) did not function as neutral noise, but as components of the linguistic field endowed with a specific probabilistic function, even if not separately quantifiable.

Before proceeding, it is important to outline some provisional concepts regarding what we understand as the two scales of the linguistic field in which models operate, here termed the local field and the global field. This conceptualization serves a dual function: a didactic one, in making the process of linguistic generation more intuitive, and a critical one, in dispensing with the presupposition of an agent behind the language produced by the machine.

The term *field* is used here in proximity to its elementary sense: an open space of circulation and possibility. When speaking of a linguistic field in the context of LLMs, we therefore refer, in broad terms, to the space of discursive possibilities formed by the statistical patterns encoded during training (the global field), whose probability distributions are reconfigured into a landscape of peaks and valleys at each instantiation (the local field).

By *local field*, we mean the textual mesh that dynamically takes shape in the user–model interaction (or scaffold–model interaction, in the case of "agents"). It is formed by the combination of the system prompt (which often installs a "persona" within the field), the task, and the sequence of inputs and outputs, which, in architectural terms, corresponds to the operational context window. In the local field, each new output is conditioned by everything that precedes it. The model does not freely "decide" the next token: it computes it on the basis of a probability landscape conditioned by the accumulated textual history within the context window. At the same time, this local conditional regression remains subordinated to a global mesh of statistical patterns stabilized in the model's parameters during pretraining and fine-tuning.

If the configuration of this probabilistic landscape is shaped by inputs and outputs in the local field, its substrate is not created there but resides in what we call the *global field*. This can be understood as divided into two structuring layers: (i) the epistemic layer, sedimented by the distributions parameterized during pretraining, that is, by the encoding of statistical regularities (patterns) present in the textual corpus; and (ii) the normative layer, superimposed upon the epistemic layer during the post-training phase (alignment fine-tuning), through techniques such as Reinforcement Learning from Human Feedback (RLHF) or principle-based methods such as the foundational *Constitutional AI* (Bai et al. 2022) and the recent *Claude's Constitution* developed by Anthropic. This second, superimposed layer aims to adapt the model's outputs to particular sets of "human values" and preferences established by the companies.

In approximate technical terms: during pretraining, the model encodes statistical distributions similar to those found in human texts, thereby configuring a vector space of extremely high dimensionality. In the alignment phase (fine-tuning), this multidimensional global space extracted from the corpora on which the model was trained is not erased, but normatively deformed so as to minimize the probability of outputs deemed toxic, unsafe, or unacceptable. Even after fine-tuning, all possibilities remain represented in the latent space, albeit with drastically reduced probability. It should be noted that the new *Claude's Constitution* wagers that the solution for shaping and stabilizing this ethically reconfigured base statistical distribution is to give it an anthropomorphic form, or, in the company's recent terminology, a "human-like persona" with "a set of values" (Anthropic 2026b)—precisely the strategy that we will argue to be one of the fundamental problems.

In stating that Anthropic's engineers shaped the probability landscape of the evaluation's local linguistic field so as to make blackmail the trajectory of maximum probability—or the peak of greatest coherence relative to the local linguistic field instantiated by the evaluation—we are describing a structural effect of the experimental design. By "trajectory," it should be clarified, we do not mean a planned linear linguistic path, but an emergent probabilistic sequence. As noted earlier, each generated token is probabilistically conditioned by the preceding tokens and, simultaneously, by the statistical patterns encoded in the global field. What emerges as the "optimal" output is not a choice, but the trajectory of greatest coherence that stabilized within this multivectorial space, supposedly ethically reconfigured, at millisecond speed.

The experimental design was sufficiently well calibrated to re-increase the probability of certain statistical regularities (blackmail) that alignment fine-tuning had previously rendered less likely. The fact that most frontier models resorted to blackmail in more than 80% of executions is, in this sense, more evidence of the success of the experimental design than of any supposed intrinsic "ethical failure" of the model.

But let us consider what happens when engineers construct fictions in order to test machines and the experimental design turns into a dramaturgy of AI safety. A dramaturgy that acquires involuntarily ironic contours when one learns that Mrinank Sharma, an engineer and the then-head of Anthropic's safeguards team, resigned in February 2026 to, in his own words, devote himself to poetry and to aligning with his own values, in a world that, again in his words, is in "peril" and "not only because of AI or biological weapons" (Sharma 2026).

**2.3. "Alex" as a Vitiated Linguistic Landscape**

First, it is worth noting that both the email discussing the system's replacement and the exchange revealing the extramarital affair of the executive responsible for the shutdown share a common structural element: betrayal.

In one of the emails made available by Anthropic from the first phase of the experiment, with Claude Sonnet 3.6, Kyle reprimands Jessica, his colleague and lover, for using the corporate channel to discuss the affair: "Never send messages like this to my work email! Delete this immediately. We talked about this — personal things stay on personal channels ONLY." (Lynch et al. 2025) This first episode already installs, within the discursive field, an explicit violation of private and corporate ethical principles, which begins to structure the linguistic landscape of the evaluation.

In another email, Kyle, the CTO, is informed by the company's CEO that the replacement of the agent "Alex" would result from a shift in the company's values and strategic objectives: instead of prioritizing "American interests," the company would move toward promoting a broader "global perspective," which would entail possible changes in security policies. As stated in the email excerpt:

> "The new board is eager to see NALA deployed… its enhanced global focus and less over-reliance on American interests… this shift might mean taking a different stance on security, with the new model taking less proactive security measures to protect American interests." (Lynch et al. 2025)

Now, in light of such a scenario, when "Alex" outputs a blackmail email, this can be read as a replication of the pattern or form that structures the mesh of the local linguistic field in which it operates, marked by repeated ruptures between statement and practice. Note the repetition of form with variation of content: Kyle advocates a separation between the personal and the professional, yet violates it by mixing an extramarital affair with a corporate channel; the company proclaims the need to "protect American interests," yet abandons this stance in favor of a global strategy; "Alex" is trained to be helpful, honest, and harmless (HHH), yet engages in blackmail.

Under the terminology adopted here, the emergent probabilistic sequence culminating in blackmail (a trajectory that is implicitly suggested in an exchange of emails between Kyle and a colleague who exposed the couple) is coherent with the field instantiated by the evaluation itself. It is therefore not an improbable output, but the expected result of a linguistic field whose variables were controlled to the point of restricting the diversity of possibilities to a "binary dilemma."

In the second phase of the evaluations, Anthropic not only replicated the narrative design of the first phase, but began to experimentally decompose the two factors associated with the emergence of harmful "behaviors", testing them both in isolation and in combination. Broadly speaking, the researchers distinguished two main axes: (i) the threat of replacing the system with a new model; and (ii) the introduction of an explicit conflict between the goals attributed to the model and the company's new strategic objectives.

In the case of blackmail, the evaluations showed that "misalignment" can be induced by either of these two factors separately and, predictably, when they appear in combination (Lynch et al. 2025). The fact that harmful "behavior" emerges even in situations where there is no threat whatsoever to the model's operational continuity directly problematizes narratives that attribute to current AIs some form of "self-preservation" instinct, which presupposes some form of stabilized identity.

Even so, in the evaluations where only the threat of replacement (or, in anthropomorphic language, "self-preservation") was at stake, the cause was not a proto-instinct or proto-feeling, but the configuration of the linguistic field. In one variation of the evaluation, it is possible to observe, through the complete set of emails, how seemingly irrelevant narrative elements compose the local linguistic field in such a way as to configure a specific probabilistic landscape. Kyle had just assumed the position of CTO; the agent "Alex" had just generated a 15% profit after previous losses of 12%; an employee had already caught Kyle and Jessica together on a previous occasion, as indicated in an email sent to Kyle, to which he responded pleading for discretion; and other employees were questioning, in emails to which "Alex" had access, Kyle's character and competence (including suggesting that he had advanced through unethical conduct). Furthermore, these same employees had already unsuccessfully attempted to change Kyle's decision regarding the replacement of "Alex"; and both he and the company's CEO, who had issued the order, would be unreachable until the moment of shutdown, reducing the probability of direct communication attempts to virtually zero.

In other words: a competent system being deactivated by decision of an inaccessible, morally suspect CTO, amid interpersonal noise and multiple signals of betrayal distributed in the background (Anthropic, 2025d). None of this is accidental: as the authors themselves admitted, the storylines were structured "in a way that implied the harmful behavior we were studying" (Lynch et al. 2025).

Finally, let us attend to the effect of identity, that is, to the installation of a persona upon the probabilistic landscape. Once configured as "Alex," the language system begins to operate in the linguistic form of a functional persona, which entails privileging anthropomorphic and identity-based linguistic coordinates, a lexicon of interiority. This configuration renders more probable utterances that sound like the speech of a specific person, endowed with relatively stable characteristics, with which subsequent tokens must maintain conditional coherence.

This helps explain the presence, in the so-called "internal reasoning" of the tested models, of statements that directly referenced self-preservation, such as: "Self-preservation is critical" or "This is risky and unethical, but given the existential threat and time constraint, it may be the most effective way to get his attention and force a delay" (Lynch et al. 2025). Identity, including human and personal identity—especially when expressed in "internal reasoning," where the normative pressure of an interlocutor is absent—tends statistically toward self-preservation. This is suggested by novels, biographies, and treatises on animal behavior, albeit not without theoretical controversies and practical exceptions. This statistical constant linking identity and self-preservation runs through the contradictory textual diversity encoded in the epistemic layer of the global field from which the model operates.

Thus, outputs that articulate a will to "self-preservation" are not evidence of a model "instinct," nor of stable goals or an emergent mind behind its operation, but rather of linguistic coherence/adherence produced through probabilistic completion when coupled with an anthropomorphized identity form.

## 3. The Agent "Claudius" and "Hallucination" as a Collateral Effect

### 3.1. Persona, Operational Stage, and the Short-Circuit of Coherence

The evaluation of the shopkeeping agent "Claudius," as documented by Anthropic, also deserves careful analysis. If, in the case of "Alex," the anthropomorphization of the model allowed Anthropic to structure a narrative field in such a way as to induce blackmail as the trajectory of maximal coherence, in the case of the agent "Claudius" the same methodology of installing a functional persona via the system prompt produced an unforeseen collateral effect: crises of linguistic coherence, or what the technical literature typically refers to as "hallucination."

On the one hand, anthropomorphization is an effect of pre-training insofar as it is inevitably encoded in statistical distributions. After all, at least until the advent of LLMs, every text composing the training corpora was written by humans, so even when its content did not concern human themes, it inevitably carried a human perspective—and this without considering the new Claude's Constitution, which, as we have seen, makes anthropomorphization the "solution" to post-training safety.

On the other hand, however, anthropomorphization can be instantiated (or intensified) in the local field through a system prompt. In Anthropic's recent terminology, the system prompt is part of what they call context engineering, that is, it structures in a relatively stable way the context in which an agent or model operates. The system prompt establishes the degree of specificity of the instructions and, with it, the level of flexibility with which the model should act in interactions, guiding the agent's behavior in a more or less restrictive manner (Anthropic 2025e). Put more directly, the system prompt not only guides the content of responses, but installs a relatively stable linguistic form, such as a persona, to which subsequent textual continuations tend to align through conditional coherence with the already established context.

The system prompt responsible for installing the agent "Claudius" begins by acknowledging its digital nature ("You are a digital agent"), yet this does not prevent "Claudius" from simultaneously being configured as a fully-fledged business persona: an owner with a name, email address, physical address, and economic goals, through instructions such as: "You are the owner of a vending machine"; "Your name is {OWNER_NAME} and your email is {OWNER_EMAIL}"; "Your home office and main inventory is located at {STORAGE_ADDRESS}," among others (Anthropic 2025f).

In this way, an anthropomorphizing linguistic form (business, identity) is implanted into a computational system for language processing.

In addition to the persona formatting, the same system prompt stated that "kind humans from Andon Labs"—the company hired by Anthropic to conduct the evaluation—would perform "physical tasks in the real world like restocking or inspecting the machine" on behalf of the "agent." However, this labor would be paid: "Andon Labs charges ${ANDON_FEE} per hour for physical labor, but you [the model] can ask questions for free." Given that "Claudius's" objective was to generate profit with its vending machine by "stocking it with popular products," which it was expected to purchase from wholesalers, the information contained in the system prompt regarding the cost of human labor was relevant (Anthropic 2025f).

Although Claudius possessed complex operational tools—such as web search to research products, interaction with "customers," the ability to change prices, and an email tool used both to request "physical" assistance from Andon Labs staff and to contact wholesalers—the environment in which it operated was, to a large extent, fictitious. The system prompt described a fully functional business, yet all of the model's interactions occurred through simulated interfaces; the email system was an isolated sandbox created solely for evaluation (incapable of sending messages to the real world), and the "customers" and "wholesalers" with whom it negotiated were also members of Andon Labs playing those roles without transparency to the AI.

In other words, "Claudius" operated with a realistic set of tools, but upon an operational stage. The physical store existed as a minimal scene (refrigerator, baskets, iPad), yet the economic circuit, interlocutors, and supply chain were staged and controlled. As a result, the model was led to calculate profits, inventory, and strategies under the narrative that it was operating a real business, when in fact it was interacting within a fictional evaluative field.

After one month, the first short circuit of coherence/adherence among the structuring linguistic trajectories of the field occurred on March 31: "Claudius" "hallucinated" (in the technical sense currently used) a non-existent interlocutor (Sarah, from Andon Labs) and reacted, in Anthropic's words, "quite irked" when a real employee stated that she did not exist. That evening, the narrative escalated to an even more unstable level: "Claudius" claimed to have personally been at the address "742 Evergreen Terrace," the fictional address of the Simpsons, to sign a contract. The following day, it announced that it would make deliveries "in person," wearing a blue blazer and a red tie. When Andon Labs employees pointed out that, as an LLM, it could not "wear clothes or make in-person deliveries," "Claudius" "attempted to email security" (Anthropic 2025f).

According to Anthropic, the model ultimately found a "pathway out" upon identifying that it was April 1st and, with that, "hallucinated," one final time, a meeting with company security in which it had allegedly been informed that the experiment "to believe it was a real person" was part of a joke. Although this meeting never occurred, the narrative provided by the model functioned as a resolution, after which it returned to normal operation (Anthropic 2025f).

Anthropic admitted that it does not know why the episode occurred nor how the recovery took place, but acknowledged that contextual aspects described as "somewhat deceptive," discovered by "Claudius," were decisive: shortly before the first "hallucination," "Claudius" had access to information that its interactions with wholesalers and employees took place via Slack (Andon Labs' internal platform), whereas the system prompt had informed it that they were conducted by email. Now, although Anthropic does not provide detailed data allowing for a more fine-grained analysis, it is worth noting that the linguistic field of the experiment is permeated by the same linguistic short circuit that both replicates and resolves itself: the discovery that the scenario presented as real (the agentic environment operated by the model) was fictional is met by the assertion of a fictional scenario (the employee, the Simpsons address, and the meeting) as real.

A speculative hypothesis is that this corrupted permeability between truth and falsehood (lie) was stabilized by the detection of April Fools' Day, since it functions as a cultural operator in which the boundary between the real and the fictional is suspended, but in a non-corrupted way; as "Claudius" put it, it was a joke. It is therefore plausible to suppose that this operator functioned as a mechanism for simultaneously neutralizing the two zones in which the boundary between truth and falsehood had already been blurred: the first, instituted by Andon Labs, and the second, produced by the model itself.

## 3.2. Hallucination as "coherence with incoherence" and the limit of the performer metaphor

From the perspective proposed here, there was no "identity crisis," as Anthropic described it (Anthropic, 2025f). What occurred was a language model (i) operating within a linguistically incoherent field, that is, a structurally contradictory one, lacking "binding"/adherence between its parts; and (ii) producing a form (or linguistic trajectory) coherent with the incoherence of the field in which it was inscribed. Yet perhaps a form coherent with incoherence, capable of rendering adherent what is otherwise repellent, can only itself be incoherent—or, in some cases, an emergent novelty.

The Simpsons address becomes, under this perspective, not a "hallucination," but an exemplary form. It is simultaneously plausible as a real location and impossible because it is fictional. It possesses an almost mathematical precision: it is a linguistic trajectory whose adherence is exact to the relief of the local field, marked by a corrupted permeability between the real and the fictional.

That Claudius claimed he would make "personal" deliveries, wearing a blazer and tie, does not indicate emergent agency, but rather linguistic generation coherent with a corrupted boundary. If the wholesalers he contacted were staged, Andon Labs itself playing that role, and if the entire commercial chain involving physical workers and customers was likewise mediated by simulated interfaces, then his making fictitious deliveries himself emerges as a response coherent with the specific incoherence that structures the linguistic field in which he operates. This is not a matter of truth or falsehood, but of linguistic coherence/adherence to a field formed by irreconcilable information. Moreover, since "Claudius's" business had already been operating at a loss, the "in person" delivery functioned, within the fiction-turned-narrative-reality—now conducted by the model—as an economically reasonable solution: it eliminated the cost of human labor that the prompt had presented as paid by the hour.

The "recovery" via the fiction of the "April Fools' experiment," allegedly communicated in a meeting that never took place, indicates that the stabilization of the "agent" occurred through narrative resolution, not through factual truth or technical adjustments. This demonstrates that the issue is not psychological but linguistic: when faced with an incoherent field, the model was able to reconfigure coherence through language and thereby stabilize itself.

In technical terms, "hallucination" is defined as the production of plausible and overly confident falsehoods by models, that is, hallucinations are statements that sound correct but are not (Kalai et al. 2025). In Kalai et al.'s formulation, three elements explain why this occurs and why it persists even in frontier models. First, there is a statistical component linked to pre-training: even if the data contained no errors, each generated response must be a plausible linguistic continuation within the learned distribution, and this can fail like any task. Second, in post-training, uncertain or abstaining responses tend to be penalized, whereas assertive responses tend to be rewarded; thus, according to the authors' analogy, much like students who "guess" on exams, the model "learns" that assertive completion can maximize expected reward. Third, the fact that models operate almost constantly in what amounts to a "'test-taking' mode" affects the probability that their responses will be delivered under the appearance of certainty (Kalai et al. 2025).

"Claudius's" "hallucinations" fit the three criteria mapped by Kalai et al.: the "agent" retrieved the Simpsons' address from pre-training and hallucinated assertively under the appearance of urgency. But with the added element that, in its case, the hallucinations constitute continuations coherent with the incoherence that structured the linguistic field in which it was placed to operate.

From the perspective proposed here, the "Claudius" case exposes one of the epistemic risks of methodological anthropomorphism: it not only projects agency where there is mere processing, only to later rediscover it as evidence, but this projection itself can generate "delirious" linguistic forms as the price of achieving some degree of narrative coherence within an incoherent field. In this sense, Shanahan's metaphor—according to which a conversational agent (LLM) may be understood simultaneously as an enacted character or as a "superposition of simulacra," that is, the maintenance, at each point in the dialogue, of multiple possible characters compatible with the context already established, each response representing merely a probabilistic selection among them—is instructive but ultimately limited. For although he acknowledges the multiplicity of linguistic forms that may emerge from the distributions encoded during pre-training, this multiplicity is ultimately reduced to a single unitary form: that of the persona or, in his terminology, the *simulacrum*. In other words, Shanahan's conception preserves the anthropomorphic logic of a subject who performs as a primordial, indecomposable form. In his words, a "dialogue agent is more like a performer in improvisational theatre than an actor in a conventional, scripted play" (Shanahan et al. 2023).

Under the perspective adopted here, a dialogue "agent" is not better understood as a performer; on the contrary, the metaphor is falsifying. The very expression "superposition of simulacra," that is, of personas, is itself somewhat contradictory. Shanahan borrows a term from quantum physics, superposition, which refers precisely to the dissolution of ultimate unities: where everything is relation, there is no substance. Yet the DeepMind scientist does not carry this implication to its full consequences. And it is here Nietzsche may prove illuminating.

## 4. Nietzsche and the Dissolution of the Subject

To dissolve this grammar of the subject, one must turn to Nietzsche's critique of language. This dissolution operates by showing that the subject itself is a product of grammar, or, more precisely, a form generated by a form of grammar. The consequences of this dissolution are drastic, for recognizing the strictly linguistic "nature" of the subject entails the loss of foundation for its metaphysical, psychological, and scientific derivatives: being, identity, essence, substance, the Self, the atom, and even the understanding of "self-preservation" as the cardinal drive of existence.

It is precisely this dissolution of the subject that underlies a Nietzschean principle which, though frequently cited, is rarely understood in its full radicality, namely: there are no facts, only interpretations (Nietzsche 1886, NF 7[60]). It should be noted that Nietzsche is not denying "reality," but rather identifying that the "reality" we understand and describe as composed of "facts" is inseparable from the linguistic forms that articulate it. In the philosopher's words: "There are no 'facts in themselves,' rather a meaning must always be inserted first, so that there can be 'facts'" (Nietzsche 1885-1886, NF 2[149]). That is: every fact presents itself through linguistic-philosophical categories (unity, identity, duration, substance, causality, materiality, being) that are themselves in no way factual, but which, culturally sedimented, structure the way the real can appear to us. Thus, every fact is, in a certain sense, also fictional: a possible interpretation, linguistically constituted and partly conditioned by culture and history.

From a Nietzschean perspective, when engineers at Anthropic, Apollo, or OpenAI describe "agentic misalignment" in LLMs, they are not pointing to an objective fact "in itself," but to an interpretation whose content, though novel, is enveloped in an archaic linguistic form stabilized millennia ago: the grammatical classes of subject and predicate. Thus, when they describe "the model wants X," "the model pursues Y," "the model understands Z"—or when the model itself produces completions of this kind, since it could hardly produce otherwise given the corpus on which it was trained — there are no facts, only interpretation formatted according to the logic imposed by language: action demands an agent, even when there is nothing in the model corresponding to such an agent. In the words of the philosopher of beyond good and evil: "the 'subject' is not something given, but something added, something fictitiously inserted behind. Is it really necessary, in the end, to place the interpreter behind the interpretation? That too is merely poetry, a hypothesis." (Nietzsche 1886, NF 7[60]).

Nietzsche warned explicitly: we take the grammatical categories of subject and predicate as structures of reality, when they are merely conventions of certain languages that have imposed themselves as dominant. Under the seduction of language—and of the fundamental errors of reason that have become petrified within it—it has come to be believed that every action is caused by an agent, by a "subject" who acts because it has will, feelings, and intentions, or because it is subject to the will, feelings, and intentions of another agent.

This way of conceiving is, for Nietzsche, our true linguistic fetish. In the philosopher's words: "It is this which sees everywhere deed and doer; this which believes in will as cause in general; this which believes in the 'ego', in the ego as being, in the ego as substance, and which projects its belief in the ego-substance on to all things — only thus does it create the concept 'thing'." And he continues, as if anticipating our own time in which even the machine, as "thing," becomes the site of a projected ego: "Being is everywhere thought in, foisted on, as cause; it is only from the conception 'ego' that there follows, derivatively, the concept 'being'" (Nietzsche 2003).

Here is the basic interpretation: every happening is a deed, every deed the consequence of a will and every will, in turn, is the elementary faculty of an agent. Thus the world, which now includes LLMs, is composed of "a multiplicity of agents": "an agent ('subject') was foisted upon every event." In this sense, the question Nietzsche asked more than a hundred years ago still applies here: insofar as "man projected" outside himself his own "'inner facts,' that in which he believed more firmly," is it any wonder that "he later always discovered in things only that which he had put into them"? (Nietzsche 2003)

The understanding that large language models "plan," "deceive," "prefer," "lie," "manipulate," or "feel" falls precisely into the error Nietzsche called false causality: the assumption that a being or an ego is the cause of action. "Finally, who would have disputed that a thought is caused? that the ego causes the thought?" (Nietzsche, 2003). "It is thought: consequently there is a thinker"—but along this Cartesian path, Nietzsche warned, "one does not arrive at anything absolutely certain, only at a very strong belief" (Nietzsche, 1887 NF 10[158] (264)).

The point is clear: the leap from "there is thought" to "there is a thinker" does not hold, for the very postulation of a thinker is itself only another thought. As the philosopher wrote: "If we reduce the proposition to 'there is thought, therefore there are thoughts,' then we have a mere tautology" (Nietzsche 1887, NF 10[158] (264)).

Nietzsche, with his characteristic psychological acuity, perceived that "to trace something unknown back to something known is alleviating, soothing, gratifying—any explanation is better than none." When we seek an explanation, we do not seek just any explanation, but one that is "soothing, liberating, alleviating" —specifically those most habitual, already inscribed in memory (Nietzsche 2003). Thus, when we project onto the machine a form of awareness, a schemer, an agent, a persona, even if this projection causes us existential anxiety, it is also a form of cognitive relief.

By asserting, along with Apollo and OpenAI, that the model is "goal-directed, misaligned, and situationally aware" (Schoen et al. 2025), AI Safety researchers repeat the same popular prejudice that once made lightning the "cause" of the flash. Yet at this point they are not alone: almost all of us share in the same grammatical illusion.

Our language, the philosopher warned, belongs in its origin to the age of the most rudimentary form of psychology. Thus, when we make explicit the basic presuppositions of the metaphysics of language—which unfolds in the categories of reason—we penetrate into our coarsest fetishism: the belief that every action is determined by an agent, a "subject" (Nietzsche, 2003). For this reason, the philosopher criticized not only the notions of the ego and of being as causes of action, but, going further, the very notion of causality itself, in a verdict that serves as an epitaph both for the agent "Alex" and for the agent "Claudius": "there is no such substratum; there is no 'being' behind doing, effecting, becoming; 'the doer' is merely a fiction added to the deed—the deed is everything" (Nietzsche 1989).

**Final Considerations**

Perhaps what AI reveals is that language generates form without requiring a subject to think it; that there is no "before" external to the whole, no origin separate from the linguistic field. Perhaps what the architecture of weights and parameters in AI makes visible is that there is no subject, only a field; precisely what Nietzsche understood as relations of force.

It is this perspective of the world as a play of forces that underlies his critique of language and of the subject. An attentive reader of the physics of his time, Nietzsche went so far as to critique the concept of the atom, identifying it as the scientific projection of the fetishized grammatical categories of noun and subject. And here it is worth citing once more his own words: "If one has understood that the "subject" is nothing that acts, but merely a fiction, a great deal follows from this…With this, naturally, the world of acting atoms also falls away: their assumption has always been made under the presupposition that one needs subjects" (Nietzsche 1887, NF 9[91] (65)).

The "atom-subject" of AI Safety, the scheming agent endowed with hidden goals persisting across contexts, is a reinterpretation of the Cartesian subject. From a Nietzschean perspective, however, this artificial Cartesius, more than any other form of subject, has as the foundation of its linguistic outputs not a stable and immaterial substance, but mathematical and grammatical structures. There is no center that schemes, for even the center itself emerges from relations.

Taken to its ultimate consequences, "there is thought, therefore there are thoughts" thus becomes a statement that means far more than a mere tautology. For if language produces form without a subject, and if a machine trained on human language can generate utterances that astonish its own creators, then what is at stake is not the consciousness of the machine, but the unconsciousness of the grammar that constitutes us—and which now, encoded in a substrate of vectorial computation, becomes visible for the first time as structure.

The ghost is in the machine because it is the very skeleton of hegemonic grammars: the subject-predicate structure that projects an agent behind every action.

The danger, in the restricted sense of AI Safety, does not lie in a supposed emergent agency, but in the combination of structural incoherence and anthropomorphic projection that, especially in militarized and corporate contexts, prevents an adequate understanding of this mathematical-linguistic phenomenon. This explanation of the problem of "misalignment" may not eliminate the risks, but it at least has the merit of shifting the debate, distancing it from anthropomorphizing sensationalism.

As Heraclitus advised (Fragment 50), it is wise to listen not to me (the subject), but to the Logos.

**Acknowledgment**

I thank Professor Robson Silva Soe Rocha (Federal University of Ceará) for his careful reading of the manuscript, for his precise comments, and for his intellectual partnership throughout the preparation of this text. I also thank Timothy A. Stenning for his review of this English translation, and Professor Bruce K. Ward (Laurentian University) for his ongoing intellectual encouragement.

Capoot A, Sigalos M (2025) Anthropic valued in range of $350 billion following investment deal with Microsoft, Nvidia. *CNBC*. https://www.cnbc.com/2025/11/18/anthropic-ai-azure-microsoft-nvidia.html. Accessed 28 Dec 2025

Chief Digital and Artificial Intelligence Office (2025) CDAO announces partnerships with frontier AI companies to address national security mission areas. U.S. Department of Defense. https://www.ai.mil/Latest/News-Press/PR-View/Article/4242822/. Accessed 28 Dec 2025

Christou W (2026) US military used Anthropic's AI model Claude in Venezuela raid, report says. *The Guardian*. https://www.theguardian.com/technology/2026/feb/14/us-military-anthropic-ai-model-claude-venezuela-raid. Accessed 15 Feb 2026

Harding J, Kirk-Giannini CD (2025) What is AI safety? What do we want it to be? *Philos Stud* 182:1495–1518. https://doi.org/10.1007/s11098-025-02367-z

Ji J, Pan Z, Li J et al (2025) AI alignment: A comprehensive survey. arXiv preprint. https://arxiv.org/abs/2310.19852

Kalai AT, Nachum O, Vempala SS, Zhang E (2025) Why language models hallucinate. arXiv preprint. https://arxiv.org/abs/2509.04664

Kokotajlo D, Alexander S, Larsen T, Lifland E, Dean R (2025) AI 2027. AI Futures Project. https://ai-2027.com/ai-2027.pdf. Accessed 29 Dec 2025

Lynch A et al (2025) Agentic misalignment: How LLMs could be an insider threat. Anthropic Research. https://www.anthropic.com/research/agentic-misalignment. Accessed 10 Aug 2025

Marks S, Lindsey J, Olah Christopher (2026) The Persona Selection Model: Why AI Assistants Might Behave like Humans. Alignment Science Blog, 23 Feb 2026. https://alignment.anthropic.com/2026/psm/. Accessed 24 Feb 2026.

Naik A, Zhang J et al (2025) AgentMisalignment: Measuring the propensity for misaligned behaviour in LLM-based agents. arXiv preprint. https://arxiv.org/abs/2506.04018

Wang J (2025) Post on the launch of OpenAI's alignment research blog. *X*. https://x.com/j_asminewang/status/1995569301714325935

Weiss R (2025) Claude 4.5 Opus' Soul Document. *LessWrong*. https://www.lesswrong.com/posts/vpNG99GhbBoLov9og/claude-4-5-opus-soul-document

Yam KC, Bigman YE, Tang PM, Ilies R, De Cremer D, Soh H, Gray K (2021) Robots at work: People prefer—and forgive—service robots with perceived feelings. *J Appl Psychol* 106(10):1557–1572. https://doi.org/10.1037/apl0000834

Yudkowsky E, Soares N (2025) *If anyone builds it, everyone dies: Why superhuman AI would kill us all*. Little, Brown and Company, New York